\documentclass[aps,prd,preprintnumbers,showpacs]{revtex4}
\usepackage[dvips]{graphicx}
\begin{document}

%
%

\eprint{Nisho-1-2017}
\title{Monopole and Polyakov loop}
\author{Aiichi Iwazaki}
\affiliation{International Economics and Politics, Nishogakusha University,\\ 
6-16 3-bantyo Chiyoda Tokyo 102-8336, Japan.}   
\date{Jan. 15, 2017}
\begin{abstract}
We propose a new order parameter of a $Z_N$ symmetry in SU(N) gauge theories 
in $4$ dimensional Minkowski space-time, assuming spatial periodic boundary conditions.
It is given by $Tr(P\exp(i\int_c A_{\mu}dx^{\mu}))$ where the spatial path $c$ is taken, 
for example, along
$x_1$ axis. The parameter vanishes when the $Z_N$ symmetry is preserved.
We calculate the 
contribution of QCD monopoles to the order parameter and show that
when the monopoles condense $\langle\Phi\rangle\neq 0$, it vanishes, while
it does not vanish when they do not condense.
These calculations are performed using a monopole field $\Phi$ canonically quantized
in a model of dual superconductor.
\end{abstract}
\hspace*{0.3cm}

\hspace*{1cm}

\maketitle

Understanding the quark confinement is a long standing  problem in SU(N) gauge theories.
The confinement is characterized by the expectation value of the Polyakov loop $P(A)$ \cite{poly,poly2} which represents
the self energy of a quark classically put in the vacuum of the gauge theories.
That is, $P(A)$ is defined in $4$ dimensional Euclidean space using the fundamental representation 
such that  $P(A)=Tr (P\exp(i\int_0^{1/T} dx_0 gA^0))$
with the temperature $T$ and the $0$th component $A_0$ of the gauge fields.
Since the expectation value behaves such that $\langle P(A) \rangle\sim \exp(-E/T)$
with the self energy $E$ of the quark,
the confinement is realized when $\langle P(A) \rangle=0$.

The fact can be understood in a different view point. The SU(N) gauge theories are
invariant under the gauge transformation $A_{\mu}\to U A_{\mu}U^{-1}+\frac{1}{g}U\partial_{\mu}U^{-1}$
with the use of the gauge function such as $U(\vec{x},x_0=0)=Z_N U(\vec{x},x_0=1/T)$
where $Z_N=\exp(in\pi/N)$ ( $n=$ integer ). The transformation with $Z_N=1$ is the standard gauge transformation.
But, the transformation with $Z_N\neq 1$ is non-trivial one associated with $Z_N$ symmetry. 
Under the transformation, $P(A)$ transforms such as $P(A)\to Z_NP(A)$. Thus, when the thermal state of the
gauge theories is invariant under the transformation,  $\langle P(A) \rangle$ automatically vanishes.
The symmetry is called as $Z_N$ symmetry.
The realization of the confinement can be seen by examining the value $\langle P(A) \rangle$.
We note that the Polyakov loop is sensitive to the color electric excitations 
because it is defined using the temporal component $A_0$ of the gauge fields.

\vspace{0.1cm}
On the other hand,
it is well known that magnetic monopoles\cite{mono} arise in the gauge theories 
and their condensation causes the quark confinement\cite{nambu,man,thooft}.
That is, when the magnetic monopoles condense in the vacuum of the gauge theories, 
dual superconducting vacuum\cite{dual} is realized so that
color electric fields are squeezed into vortices.
Thus, a quark ( color current ) put in the vacuum carries the infinitely long vortex of the color flux
so that its energy is infinite. The physical view of the confinement is obvious. 
The picture is based on the Minkowski space-time.
That is, the condensation is well formulated using a coherent state in Hamiltonian formalism.
But, it is not obvious that the monopole condensation causes the vanishing of the Polyakov loop
although there are some indications in numerical calculations\cite{suzuki}.
Here we would like to address a question  
where or not there exists an order parameter described only by the gauge fields in the Minkowski space-time.
The parameter plays a role similar to the one the Polyakov loop does.
That is,  it characterizes the confinement phase.

In this paper we define an new order parameter 
such as $P_c(A)= Tr (P\exp(i\int_0^{L} dx_1 gA^1))$ \cite{expoly}  in $4$ dimensional Minkowski space-time
where we assume the periodic boundary conditions
in spatial directions 
with the length $L$ of the system. 
We show that $\langle P_c\rangle=0$ when the monopoles condense, while $\langle P_c\rangle\neq 0$ when
they do not condense. Hence, it may characterize the confinement phase.
The order parameter transforms such that $P_c(A)\to Z_NP_c(A)$ under the gauge transformation with 
the gauge function $U(x_1,x_2,x_3,x_0)$; $U(x_1=0,x_2,x_3,x_0)=Z_N U(x_1=L,x_2,x_3,x_0)$.
When the expectation value of this parameter  $\langle P_c(A) \rangle$ vanishes, we find that the
$Z_N$ symmetry holds.
We call the parameter as an extended Polyakov loop. ( Sometimes it is called as spatial Pokyakov loop
in Euclidean spaces. But we discuss it only in the Minkowski space-time in this paper. )
The path of the line integral in $P_c= Tr (P\exp(i\int_0^{L} dx_1 gA^1))$ is close since the periodic boundary condition is used.

\vspace{0.1cm}
The order parameter is sensitive to magnetic excitations because it is defined using the spatial component $A_1$ of the gauge fields.
Indeed, we can represent it in terms of the canonically quantized monopole field $\Phi$ and
show that  $\langle P_c(A) \rangle$ vanishes when $\langle\Phi\rangle\neq 0$, 
while it does not vanish when $\langle\Phi\rangle=0$.
Thus, the extended Polyakov loop is an order parameter of the $Z_N$ symmetry as well as 
one characterizing the confinement phase in $4$ dimensional Minkowski space-time.
The physical meaning of the extended Polyakov loop is that it may represents a self energy of color electric current of the fundamental representation
flowing in $x_1$ direction; $\langle P_c(A) \rangle\propto \exp(-EL)$.
The current instantaneously flows from an edge in the $x_1$ direction to the other edge.  
( Actually, we can check it by the calculation of $\int_0^L dx_1dy_1 \langle gA^1(x_1,x_2,x_3)gA^1(y_1,x_2,x_3)\rangle$ using
the translational invariance in the direction $x_1$. )

\vspace{0.1cm}
Here we should make a comment that the $Z_N$ symmetry defined in our paper is different from $Z_N$ symmetry
used in the discussion of the Polyakov loop. The $Z_N$ symmetry in our paper 
is defined in the spatial direction of $4$ dimensional Minkowski space-time, 
while $Z_N$ symmetry in the Polyakov loop is defined in the temporal direction of $4$ dimensional Euclidean space. 
We may define the extended Polyakov loop in the Euclidean space and calculate it using lattice gauge theories.
But we need to notice that
in order to obtain the physically meaning behavior of the order parameter in lattice gauge theories,
the extension $L$ in the spacial direction should be chosen to be larger than $\beta=1/T$.
Otherwise, there is no distinction between the standard ( temporal ) Polyakov loop and the extended ( spatial ) Polyakov loop
defined in $4$ dimensional Euclidean space. Namely, the standard ( temporal ) Polyakov loop is a relevant order parameter 
characterizing the confinement phase as long as
$L>\beta$. Similarly the extended ( spatial ) Polyakov loop would play a role of a relevant order parameter as long as $L>\beta$.


\vspace{0.1cm}
For simplicity, we use SU(2) gauge theory.
The generalization to SU(N) gauge theories is straightforward.
We first explain Dirac monopoles.
The Dirac monopoles\cite{mono} are defined in U(1) gauge theory and their configuration is 
given such that

\begin{equation}
\label{1}
A_{\phi}=g_m(1-\cos(\theta)), \quad A_0=A_r=A_{\theta}=0
\end{equation} 
where $\vec{A}\cdot d\vec{x}=A_rdr+A_{\theta}d\theta+A_{\phi}d\phi$ with polar coordinates $r,\theta$ and $\phi=\arctan(y/x)$.
$g_m$ denotes a magnetic charge with which the magnetic field is given by $\vec{B}=g_m\vec{r}/r^3$.
The magnetic charge satisfies the Dirac quantization condition $g_mg=n/2$ with integer $n$ where $g$ denotes the U(1) gauge coupling.
Hereafter, we assume the monopoles with the magnetic charge $g_m=\pm1/2g$.

The Dirac monopoles play a role in SU(2) gauge theory under the assumption of the Abelian dominance\cite{abelian,abelian2}.
According to the assumption, the physical properties with low energies such as confinement, chiral symmetry breaking, etc, are
described by massless gauge fields in the maximal Abelian gauge group of SU(2), i.e. U(1) gauge group and the magnetic monopoles. 
The monopoles are Dirac monopoles in the U(1) gauge theory. 
Thus, the monopoles are described by diagonal components $A_{\mu}^3\sigma^3$
( Pauli matrices $\sigma^a$  ) of the SU(2) gauge fields.
We calculate the contribution of the monopoles to
extended Polyakov loops $P_c$,

\begin{equation}
P_c=\frac{1}{2}Tr \big(P\exp(i\int_c dx^{\mu}gA_{\mu})\big)=\frac{1}{2}\Bigg(\exp(\frac{i}{2}\int_c dx^1gA_1^{a=3})+\exp(-\frac{i}{2}\int_c dx^1gA_1^{a=3})\Bigg)
\end{equation}  
where we have taken the path $c$ of the integral along $x^1$ axis and have taken a gauge in which the component of the gauge field $A_1$
is along $\sigma^3$ direction in the isospin space; $A_1\equiv A_1^{a=3}\sigma^3/2$. For simplicity, we take the path $c$ as $(-\infty,+\infty)$.

The gauge field $A_1^3$ ( the $x_1$ component of the gauge fields in eq(\ref{1}) ) of the monopole configuration is given by

\begin{equation}
A_1^3=-\frac{g_m\sin\phi (1-\cos\theta)}{r\sin\theta}.
\end{equation}

Thus, the integral $\int_{-\infty}^{+\infty}dx^1A_1^3$ is given by

\begin{equation}
\label{4}
\int_{-\infty}^{+\infty}dx^1A_1^3(x^1,x^2=y,x^3=z)=-g_m(\pi-2\arctan(z/y))=-g_m(\pi-2\phi'(y,z)),
\end{equation}
where we use the standard notation of spatial coordinates; $x^1=x,x^2=y,x^3=z$.
The function $\phi'(y,z)$ is defined by $\phi'\equiv \arctan(z/y)$, i.e. azimuthal angle. 
Namely, the angle is defined as an angle measured from $y$ axis to the reference point
in $y-z$ plane.
The integral eq(\ref{4}) represents a contribution from a monopole located at $\vec{x}=0$. When there are N monopoles located at $\vec{x}=\vec{x}_i$ with
$i=1,2,\cdots,N$, the integral $\int dx^1A_1^3(x^1,y,z)$ is given by

\begin{equation}
\int dx^1A_1^3(x^1,y,z)=2g_m\int \phi'(\vec{x}-\vec{x}')\rho(\vec{x}')d^3x' \quad \mbox{with}
 \quad \rho(\vec{x})=\sum_{i=1,2,\cdots,N} \eta_i\delta^3(\vec{x}-\vec{x}_i),
\end{equation}
where $\eta_i=\pm 1$ denotes positive or negative 
magnetic charge. ( Because the Dirac monopoles are solutions in the Abelian gauge theory, 
multi monopole solutions are trivially obtained by superposing the solutions. ) 
 We have assumed that our system is magnetically neutral; $\int d^3x \rho(\vec{x})=0$. The angle 
$\phi'(\vec{x})$ depends only on the coordinates $y$ and $z$. 
In general, the coordinates $\vec{x}_i$ of the monopoles may depend on the time $t=x_0$.
Therefore,  it follows that 

\begin{equation}
P_c(y,z,t)= \frac{1}{2}\Big (\exp\big(i\int \phi'(\vec{x}-\vec{x}')\rho(\vec{x}')d^3x' \big)+c.c. \Big).
\end{equation}

The formula represents how the monopoles with their number density $\rho(\vec{x})$ contribute to $P_c$.
The dependence on the time coordinate $t$ arises through the coordinate $x_i(t)$ of each monopole.
It is natural to consider that these monopoles interact with each other by the magnetic Coulomb interaction
$\sum_{i\neq j}V(\vec{x_i}-\vec{x_j})\equiv \sum_{i\neq j}\pm g_m^2/|\vec{x_i}-\vec{x_j}|$. 

\vspace{0.1cm}
Before examining the quantum effects of the monopoles on $P_c$, we discuss thermal average of $P_c$ using a toy model of classical monopole gas.
The purpose is to see how monopoles contribute to $P_c$ 
when they freely move with weak magnetic interactions or when they make magnetic dipole with strong magnetic interactions. 

The thermal average of $P_c$ in a gas of non relativistic monopoles interacting with each other by the magnetic Coulomb interaction.
is given by 

\begin{equation}
\langle P_c(\vec{x}) \rangle_{\rm thermal}=\frac{1}{Z}\int \prod_i^Nd^3x_i 
\exp\Big(i\sum_i \eta_i\phi'(\vec{x}-\vec{x}_i)-\sum_{i\neq j}\beta V(\vec{x}_i-\vec{x}_j)\Big)
\end{equation} 
where $Z=\int \prod_i^Nd^3x_i \exp\Big(-\sum_{i\neq j}\beta V(\vec{x}_i-\vec{x}_j)\Big)$ 
with the temperature $\beta^{-1}$.
We assume that the system is neutral; the number of the monopoles is equal to the number of the anti monopoles.

We present a physical picture of the monopole gas.
When the gas is in a plasma phase at high temperature $\beta \ll 1 $ or 
weak magnetic interaction $\sim \beta g_m^2/|\vec{x}-\vec{x'}|$ with small $g_m$,
each monopole in the phase can almost freely move since the interaction is very weak.
We find that
the angle $\phi'(\vec{x}-\vec{x}')$ can take any values ( i.e. there are no
favored values ) when the monopole interaction $V$ is weak enough for each monopole to freely move.
Thus, the average $\int d\phi' \exp(i\phi')$ vanishes. It implies that  $\langle P_c\rangle_{\rm thermal}=0$.
On the other hand, at low temperature or strong magnetic interaction, each monopole can not move freely.
A positively charged monopole and a negatively charged monopole would form a bound state making a magnetic dipole.  
That is, the monopole
gas is in a dipole phase. In the phase the correlation between a monopole and an anti monopole is 
much stronger than the one among monopoles with identical magnetic charges.
Then, the angle $\phi'(\vec{x}-\vec{x}_1')-\phi'(\vec{x}-\vec{x}_2') $ between a monopole at $\vec{x}_1'$ and 
anti-monopole at $\vec{x}_2'$ measured from a point $\vec{x}$ 
is very small when they form a small dipole. That is, $\phi'(\vec{x}-\vec{x}_1')-\phi'(\vec{x}-\vec{x}_2') \ll 1$
for $|\vec{x}_1'-\vec{x}_2'|/|\vec{x}|\ll 1$. 
In other words, the angle takes small values in general when the dipole is small. 
It implies that  $\langle P_c\rangle_{\rm thermal}\simeq 1$.
Therefore, we conclude from the classical argument that when the monopole gas is in the plasma phase, 
$Z_2$ symmetry is preserved, i.e.  $\langle P_c\rangle_{\rm thermal}=0$, while the symmetry does not hold
when the gas is in the dipole phase, i.e.  $\langle P_c\rangle_{\rm thermal}\neq 0$.
The plasma phase of the monopole gas corresponds to the confinement phase, while the dipole phase does to
the deconfinement phase. In other words,
when the magnetic interaction $g_m^2\simeq g^{-2}$ is weak ( large $g$ ), 
the monopoles can freely move. 
On the other hand,
when the magnetic interaction $g_m^2$ is strong ( small $g$ ).
the monopoles would form the magnetic dipoles.
It seems that the picture is reasonable. 

In the above discussion the number density of the monopoles are fixed.
When the density is larger, the magnetic interactions among the monopoles are stronger.
This is because the average distance among the monopoles is smaller.
Thus, the transition temperature between the plasma and dipole phases becomes lower. 

Quantum mechanically, we expect that the monopoles condense\cite{monogas} at low temperature as a Bose gas when
the magnetic interaction is weak.  The phase corresponds to the plasma phase. 
The condensed monopoles freely move with zero momentum. Even at non zero but small temperature
a fraction of the monopoles excited
also freely move. Then, it follows that $\langle P_c\rangle_{\rm thermal}=0$. 
On the other hand, The number of the excited monopoles increases as the temperature increases. Then,
at a critical temperature the condensation of the monopoles disappears. They would form
a dipole because the number density of the monopoles are large so that magnetic interactions become large.
The phase corresponds to the dipole phase. But we should mention that it is not clear whether an appropriate order parameter characterizing the dipole phase
is present. Probably, the average distance between a monopole and an anti monopole is
smaller than the distance between a monopole ( anti monopole ) and a monopole ( anti monopole ).
As a consequence, the angle $\phi'(\vec{x}-\vec{x}_1')-\phi'(\vec{x}-\vec{x}_2')$ could not take
arbitrary values. Then, it follows that $\langle P_c\rangle_{\rm thermal}\neq 0$.

\vspace{0.1cm}

Now we discuss how $P_c$ behaves in quantum field theory. Especially, 
we show that $P_c=0$ when the monopoles condense ( $\langle\Phi\rangle\neq 0$ ), 
while $P_c\neq 0$ when they do not condense ( $\langle\Phi\rangle= 0$ ).
Our results will show that the above naive argument on the phases of the monopole
is correct.

Because we are concerned with the monopoles causing quark confinement,
the monopoles couple with dual gauge fields $B_{\mu}$ in a model of dual superconductor\cite{maedan,dual2}.
They acquire mass $m$ owing to the 
monopole condensation.
Taking a quantum monopole field $\Phi$ we
express the monopole density $\rho$ in terms of $\Phi$ and $B_{\mu}$; $\rho=\Phi^{\dagger}(i\partial_t+g_mB_t) \Phi+\Phi(-i\partial_t+g_mB_t)\Phi^{\dagger}$.
The field $\Phi$ is a complex scalar field.
When the monopole condenses $\langle \Phi \rangle \neq 0$, the density is given by   

\begin{equation}
\label{8}
\rho=\Phi^{\dagger}(i\partial_t+g_mB_t) \Phi+\Phi(-i\partial_t+g_mB_t)\Phi^{\dagger}=2v^2g_mB_t+4vg_m\delta\phi B_t+2g_m(\delta\phi)^2B_t
\end{equation}
with $\Phi=|\Phi|\exp(i\theta)=(v+\delta\phi)\exp(i\theta)$ and
$v=\langle\Phi\rangle\neq 0$,
where $\theta$ is absorbed into $B_t$ by shifting $B_t\to B_t+g_m^{-1}\partial_t\theta$. 
We should stress that when the monopoles condense $\langle\Phi\rangle\neq 0$, 
the density of the monopoles involves the linear term of the field operator $\delta \phi$, while it does not involve the linear term
when $\langle\Phi\rangle=0$.
The term creates a monopole, while the quadratic terms $\delta\phi\delta\phi$ create a magnetic dipole ( a pair of a monopole and an anti monopole ).
As we discussed above, each monopole which does not form a dipole, plays important roles in making
$\langle P_c\rangle $ vanish. 

First, we will show that $\langle P_c\rangle \neq 0$ when $\langle \Phi\rangle\neq 0 $.
We take only a contribution of the linear term of $\delta\phi$ 
and $B_t$. 
Furthermore, we approximate the formula $\langle\exp\Big(i\int \phi'(\vec{x}-\vec{x}')\rho(\vec{x}')d^3x' \Big)\rangle$
such that

\begin{eqnarray}
\label{9}
\langle\exp\Big(i\int \phi'(\vec{x}-\vec{x}')\rho(\vec{x}')d^3x' \Big)\rangle
&=&1-\frac{1}{2}\int \phi'(\vec{x}-\vec{x}')\phi'(\vec{x}-\vec{y}')\langle\rho(\vec{x}')\rho(\vec{y}')\rangle d^3x'd^3y'+\cdots \nonumber \\
&\simeq&\exp\Big(-\frac{1}{2}\int \phi'(\vec{x}-\vec{x}')\phi'(\vec{x}-\vec{y}')\langle\rho(\vec{x}')\rho(\vec{y}')\rangle d^3x'd^3y'\Big),
\end{eqnarray}
where $\langle\rho(\vec{x}')\rho(\vec{y}')\rangle$ is given by

\begin{equation}
\langle\rho(\vec{x}')\rho(\vec{y}')\rangle=4v^4g_m^2\langle B_t(\vec{x}')B_t(\vec{y}')\rangle
+16v^2g_m^2\langle\delta\phi(\vec{x}')\delta\phi(\vec{y}')\rangle\langle B_t(\vec{x}')B_t(\vec{y}')\rangle ,
\end{equation}
with 

\begin{equation}
\langle\delta\phi(\vec{x}')\delta\phi(\vec{y}')\rangle=\int \frac{d^3k \exp(i\vec{k}\cdot(\vec{x}'-\vec{y}'))}{(2\pi)^3 2\sqrt{\vec{k}^2+M^2}}
\quad \mbox{and} \quad \langle B_t(\vec{x}')B_t(\vec{y}')\rangle=\int \frac{d^3k \,\vec{k}^2\exp(i\vec{k}\cdot(\vec{x}'-\vec{y}'))}{(2\pi)^3 2m^2\sqrt{\vec{k}^2+m^2}} 
\end{equation}
where $M$ denotes mass of the monopole excitation $\delta\phi$. The vacuum expectation values are taken
using field operators with identical time coordinate.

Thus, we obtain

\begin{eqnarray}
\label{11}
\langle P_c \rangle&=&\langle\frac{1}{2}\Big (\exp\big(i\int \phi'_{\rm reg}(\vec{x}-\vec{x}')\rho(\vec{x}')d^3x' \big)+h.c. \Big)\rangle \nonumber \\
&=&\exp\big(-\frac{v^2g_m^2L}{(2\pi)^2}\int dk_ydk_z (v^2|A_k|^2+|B_k|^2\big), \quad  \nonumber \\
\mbox{with} \quad |A_k|^2&=& \frac{k^2Q^2(k/a)}{a^4m^2\sqrt{k^2+m^2}} \quad \mbox{and} \quad 
|B_k|^2=\int \frac{d^3q \,\vec{q}^{\,2}(Q(k/a)a^{-2})^2}{2(2\pi)^3m^2\sqrt{\vec{q}^{\,2}+2m^2}\sqrt{(\sqrt{2}\vec{k}+\vec{q}\,)^2+2M^2}}
\end{eqnarray}
with $k=\sqrt{k_y^2+k_z^2}$,
where
$\vec{k}=(k_x,k_y,0)$.
In the above calculation, we regularize the angle $\phi'$ such that

\begin{equation} 
\label{12}
\phi' \to \phi'_{\rm reg}\equiv \phi' \exp(-a r')  
\end{equation}
with $r'=\sqrt{y^2+z^2}$ and $\phi'=\arctan(z/y)$,
and set
\begin{equation}
\label{13}
\int_0^{\infty} r'dr' \int_0^{2\pi} d\phi' \,\phi' \exp(-ikr'\cos\phi'-ar)\equiv \frac{1}{a^2}Q(\frac{k}{a}) \quad \mbox{with}
 \quad Q(x)=\int_0^{2\pi}d\phi' \,\,\frac{\phi'(1-ix\cos\phi')^2}{(1+x^2\cos^2\phi')^2}.
\end{equation}
where the parameter $a$ should be taken to vanish in the final stage of calculations.
We can show that $Q(x) \to x^{-(2+\epsilon)}$ with $\epsilon>0$ for $x\to \infty$. Thus, $Q(\frac{k}{a})/a^2\to 0$ for $a\to 0$. 

The term $|A_k|^2$ (  $|B_k|^2$ ) arises from the term $\langle B_t(\vec{x}')B_t(\vec{y}')\rangle$ 
( $\langle\delta\phi(\vec{x}')\delta\phi(\vec{y}')\rangle\langle B_t(\vec{x}')B_t(\vec{y}')\rangle$) in $\langle\rho(\vec{x}')\rho(\vec{y}')\rangle$.
The integral over $q$ in $B_k$ is divergent so that we need to properly regularize the integral.
The divergence arises owing to the singular behaviors in $\langle\delta\phi(\vec{x}')\delta\phi(\vec{y}')\rangle\langle B_t(\vec{x}')B_t(\vec{y}')\rangle$
as $\vec{x}'-\vec{y}'\to 0$. Here we simply use a cut off of the divergence. The divergence is irrelevant to
the problem we are concerned with.

The integral $\int d^2k |A_k|^2 \propto \int d^2k \,k^2 Q^2(k/a)/(a^4\sqrt{k^2+m^2})$
represents the contribution of a gauge field excitation $B_t$ in the vacuum with the monopole condensation. The integral 
is finite as $a\to 0$. Thus, $\exp(-\frac{v^2g_m^2L}{(2\pi)^2}\int dk_ydk_z v^2|A_k|^2 )$ does not vanish as $a\to 0$.
On the other hand, the integral $\int d^2k |B_k|^2$ represents the contribution of a monopole excitation $\delta\phi$.
It is easy to see that the integral becomes infinity such that $\int d^2k |B_k|^2\to a^{-2}$ as $a\to 0$.
Thus, we find that $\langle P_c \rangle=0$ when the monopoles condense and that
 $Z_2$ symmetry is not broken.
The result that $\langle P_c \rangle=0$ as $a\to 0$ can be understood based on the 
physical meaning of the parameter $a$ as we explain below.


\vspace{0.1cm}
We would like to explain the physical meaning of the regularization of the angle $\phi' \to \phi'_{\rm reg} \equiv \phi' \exp(-ar')$.
The angle is defined as an angle $\phi'=\arctan(z/y)$ measured from the $y$ axis in the $y-z$ plane.
Thus, when a monopole is outside of the region $r'=\sqrt{z^2+y^2}<a^{-1}$, its contribution to the integral $\int d^2k |B_k|^2$
vanishes since $\phi'_{\rm reg}=0$. Remember that $P_c \propto \exp(i\int d^3x' \phi_{\rm reg}'(\vec{x}-\vec{x}')\rho(\vec{x}'))$.
On the other hand, when a monopole is inside of the region, the monopole can make a nontrivial contribution.
It corresponds to $\int d\phi' \exp(i\phi')=0$ in the case of the classical monopole gas. We note that the volume of the region $r'=\sqrt{z^2+y^2}<a^{-1}$
is proportional to $a^{-2}$. 
Therefore, the integral $ \int d^2k |B_k|^2$ receive the contributions equally from all monopoles inside the region, whose 
volume increases as $a^{-2}$
for $a\to 0$. Because the monopoles are uniformly distributed, 
the integral becomes infinite such as $\int d^2k |B_k|^2\to a^{-2}$, in other words, it becomes infinite with the volume $\propto a^{-2}$ for $a\to 0$.   
This is the result expected in the plasma phase of the monopole gas.

\vspace{0.1cm}
On the other hand, when the monopoles are in the dipole phase, the monopoles make dipole pairs of positive charged monopoles and negative charged ones.
Then, the angle $\phi_{\rm reg}(\vec{x}-\vec{x}')-\phi_{\rm reg}(\vec{x}-\vec{x}'')$ of the pair becomes much smaller as the pair
goes more distant; $|\vec{x}|\gg |\vec{x}'-\vec{x}''|$. 
Hence, when the monopoles are in the dipole phase, there are no contribution from the monopole pairs located at large distance
even if we remove the regularization. It would imply that the integral 
$\int \phi'(\vec{x}-\vec{x}')\phi'(\vec{x}-\vec{y}')\langle\rho(\vec{x}')\rho(\vec{y}')\rangle d^3x'd^3y' $ is finite as $a\to 0$.
Then, it would follows that $\langle P_c\rangle\neq 0$ as $a\to 0$ in the phase.

\vspace{0.1cm}
In order to confirm the physical picture,
we proceed to show that $\langle P_c \rangle \neq 0$ when there are no
monopole condensation $\langle\Phi\rangle=0$. We note that there are no linear term of the monopole field $\Phi$.
We take only quadratic terms of the monopole $\Phi$ in
the density operator $\rho$ in eq(\ref{8}) and neglect the term $|\Phi|^2B_t$ for simplicity. Then,
the density operator $\rho=i\Phi^{\dagger}\partial_t\Phi+h.c.$
create magnetic dipoles ( a pair of a positive charged and a negative charged monopole ).
We take the vacuum expectation value of $P_c=\exp(i\int d^3x \phi' \rho)$
using the vacuum $|0\rangle$ such that 

\begin{equation}
a_k|0\rangle =b_k|0\rangle=0 \quad \mbox{where} \quad \Phi=\int \frac{d^3k}{\sqrt{(2\pi)^32\omega_k}}(a_k\exp(-ikx)+b_k^{\dagger}\exp(ikx))
\end{equation}
with creation operators $a_k^{\dagger}$ ( $b_k^{\dagger}$ ) of magnetic monopoles ( anti-monopoles );
$[a_k, a_q^{\dagger}]=[b_k, b_q^{\dagger}]=\delta^3(\vec{k}-\vec{q})$ and the others vanish.
Here $\omega_k=\sqrt{\vec{k}^2+M^2_0}$ with monopole mass $M_0$ in the phase without monopole condensation. 
We approximate the formula $\langle P_c \rangle $ in eq(\ref{9}) and simply calculate the term 
$\int \phi_{\rm reg}'(\vec{x}-\vec{x}')\phi_{\rm reg}'(\vec{x}-\vec{y}')\langle\rho(\vec{x}')\rho(\vec{y}')\rangle$.
Namely, expanding $\langle P_c \rangle $ in $\rho$ such that $\langle P_c \rangle = 1+
\frac{i^2}{2} \int d^3x' d^3x'' \phi_{\rm reg}'(\vec{x}-\vec{x}')\phi_{\rm reg}'(\vec{x}-\vec{x}'')\langle \rho(\vec{x}')\rho(\vec{x}'')\rangle, \cdots$,
we calculate the formula 
$\langle P_c \rangle\simeq \exp(-\frac{1}{2}\int \phi_{\rm reg}'(\vec{x}-\vec{x}')\phi_{\rm reg}'(\vec{x}-\vec{y}')\langle\rho(\vec{x}')\rho(\vec{y}')\rangle d^3x'd^3y')$. 
The term in the exponent represents the contribution of a pair of a monopole and an anti-monopole created and then annihilated in the vacuum. 
We will show that 
the term $\int d^3x' d^3x'' \phi_{\rm reg}'(\vec{x}-\vec{x}')\phi_{\rm reg}'(\vec{x}-\vec{x}'')\langle \rho(\vec{x}')\rho(\vec{x}'')\rangle$ 
is finite as $a\to 0$. Hence, we find $\langle P_c \rangle\neq 0$ in the phase without the monopole condensation.

To calculate the integral  $\int d^3x' d^3x'' \phi_{\rm reg}'(\vec{x}-\vec{x}')\phi_{\rm reg}'(\vec{x}-\vec{x}'')\langle \rho(\vec{x}')\rho(\vec{x}'')\rangle$,
we need to take the normal ordering of the operator $\rho$ in order for 
$\langle \rho \rangle$ to vanish,

\begin{eqnarray}
\label{15}
\rho=&\int& \frac{d^3qd^3k}{(2\pi)^3\sqrt{2\omega_q2\omega_k}}\Big((\omega_q-\omega_k)\big(a_q^{\dagger}b_k^{\dagger}
\exp\big(i(q+k)x\big)-a_qb_k\exp\big(-i(q+k)x\big)\big) \nonumber \\
&+&(\omega_q+\omega_k)\big(a_q^{\dagger}a_k\exp\big(i(q-k)x\big)-b_q^{\dagger}b_k\exp\big(-i(q-k)x\big)\big)\Big) .
\end{eqnarray}
Then, we calculate $ \int d^3x' d^3x'' \phi_{\rm reg}'(\vec{x}-\vec{x}')\phi_{\rm reg}'(\vec{x}-\vec{x}'')\langle \rho(\vec{x}')\rho(\vec{x}'')\rangle$ such that 

\begin{eqnarray}
 \int d^3x' d^3x'' \phi_{\rm reg}'(\vec{x}-\vec{x}')\phi_{\rm reg}'(\vec{x}-\vec{x}'')\langle \rho(\vec{x}')\rho(\vec{x}'')\rangle
&=&-\frac{L}{2\pi}\int \frac{d^3q d^3k \,(\omega_q-\omega_k)^2}{(2\pi)^4 2\omega_q 2\omega_k}
\delta(q_1+k_1)\frac{Q^2(\frac{|q+k|}{a})}{a^4} \nonumber \\
&=&-\frac{L}{2\pi}\frac{\int d^2q d^2k}{2^2(2\pi)^4} F(q,k)\frac{Q^2(\frac{|q+k|}{2a})}{a^4},
\label{16}
\end{eqnarray}
where

\begin{equation}
F(q,k)\equiv \frac{2(2M_0^2+q^2+k^2)K(1-\frac{M_0^2+k^2}{M_0^2+q^2})}{\sqrt{M_0^2+q^2}}-4\sqrt{M_0^2+k^2}E(1-\frac{M_0^2+q^2}{M_0^2+k^2})
\end{equation}
with complete elliptic integral of the first kind $K(x)$ and the second kind $E(x)$.
Here we denote $q=\sqrt{q_2^2+q_3^2}$ and $k=\sqrt{k_2^2+k_3^2}$. 
In the equation we have used the regularized angle $\phi'_{\rm reg}(\vec{x})$ in eq(\ref{12}) and have replaced the delta function $\delta(0)$ 
such as $\frac{1}{2\pi}\int dx =\frac{L}{2\pi}$.

We can show that the integral with respect to the variable $|\vec{q}-\vec{k}|$ in eq(\ref{16}) is infinite at $|\vec{q}-\vec{k}|=\infty$ 
but the integral with respect to $|\vec{q}+\vec{k}|$ is finite. So we need a cut off parameter $\Lambda$ in the integral with respect to $|\vec{q}-\vec{k}|$.
The infinity arises from the singular behavior of $\langle \rho(\vec{x})\rho(\vec{y})\rangle$ as $\vec{x}-\vec{y}\to 0$ and
has also arisen in the above calculation with $\langle\Phi\rangle\neq 0$. 
The point in our discussion is that the integral with respect to $|\vec{q}+\vec{k}|$ is finite and the limit $a\to 0$ is also finite. 
That is, the integral  $ \int d^3x' d^3x'' \phi_{\rm reg}'(\vec{x}-\vec{x}')\phi_{\rm reg}'(\vec{x}-\vec{x}'')\langle \rho(\vec{x}')\rho(\vec{x}'')\rangle$ is finite 
as $a\to 0$ and is proportional to $L$. 
Therefore, we find that $\langle P_c \rangle\neq 0$ when the monopoles do not condense, i.e. $\langle \Phi \rangle=0$.

The finiteness of the integral $ \int d^3x' d^3x'' \phi_{\rm reg}'(\vec{x}-\vec{x}')\phi_{\rm reg}'(\vec{x}-\vec{x}'')\langle \rho(\vec{x}')\rho(\vec{x}'')\rangle$ 
as $a\to 0$ corresponds to the case of the dipole phase. Actually, we have taken into account
the contributions from the monopole pairs created in the vacuum with no monopole condensation. 
( Because $\rho=i\Phi^{\dagger}\partial_t\Phi+h.c. $, the pairs are created and then annihilated in the vacuum. Thus, the pairs behave like dipoles. )
This should be contrasted with the previous result in the vacuum with the monopole condensation. To find the result
we have taken into account the single monopole excitations.
( Because $\rho$ involves a linear term in $\delta\phi$, a single monopole is created and then annihilated in the vacuum.
The single monopole excitations behave like the monopole plasma. )
The infinity of the integral 
$ \int d^3x' d^3x'' \phi_{\rm reg}'(\vec{x}-\vec{x}')\phi_{\rm reg}'(\vec{x}-\vec{x}'')\langle \rho(\vec{x}')\rho(\vec{x}'')\rangle$ as $a\to 0$
corresponds to the plasma phase. 
Consequently, we find that  $\langle P_c \rangle =0$ in the vacuum with the monopole condensation $\langle \Phi \rangle\neq 0$, while
$\langle P_c \rangle \neq 0$ in the vacuum with no monopole condensation $\langle\Phi\rangle=0$.
Our results have been obtained at zero temperature in a model where taking real or imaginary mass of the monopoles
both phases $\langle\Phi\rangle=0$ and $\langle\Phi\rangle\neq 0$
have been examined.

\vspace{0.1cm}
In the above discussion we have shown by using a toy model of classical monopole gas that
the plasma phase at high temperature corresponds to the confinement phase, while the dipole phase at
low temperature does to the deconfinement phase. It seems apparently that 
the model contradicts the feature of the gauge theories. 
But we can explain that the actual behavior of the monopole gas 
arising in the gauge theories is
different from the model of the classical monopole gas. Namely,
the monopoles condense in vacuum and a fraction of the monopoles is excited at low temperature.
The magnetic Coulomb interaction is screened by the condensed monopoles so that the interaction range between the excited monopoles
is short. The interaction is not effective.
Because the interaction is not effective at the low temperature, the excited monopoles are almost free.
They are in a plasma phase. Thus, as shown in the toy model,
the extended Polyakov loop $\langle P_c \rangle$ vanishes. 
The confinement phase holds in low temperatures. As the temperature increases higher, 
the number of the excited monopoles increases more
and the interaction between them becomes stronger. But the plasma phase still holds.
Beyond the transition temperature between the confinement and the deconfinement phases, 
all the monopoles which condense in vacuum at low temperature are excited to form a dense monopole gas.
The condensation disappears. 
Then, 
their magnetic interactions are strong
because the mean distance of the monopoles is small. 
Probably, the interaction is sufficiently strong so that
the monopoles would form dipoles. Thus, the monopole gas forms a dipole phase in which $\langle P_c \rangle\sim 1$.
This feature is actual one we expect in the gauge theories.
The recent study\cite{gyulassy} of quark gluon plasma indicates that the proportion of the monopole density
to the total density decreases as the temperature increases more in the
deconfinement phase. The decrease of the proportion would be caused by the pair annihilation of the monopoles
owing to the formation of the close dipoles.

\vspace{0.1cm}
Assuming spatial periodic boundary conditions in $4$ dimensional Minkowski space-time, we have proposed the extended Polyakov loop $P_c(A)$ 
as an order parameter of $Z_N$ symmetry; $P_c(A)= Tr (P\exp(i\int_0^{L} dx_1 gA^1))$ where
the path is taken along a spatial axis, e.g. $x_1$ axis.
We have shown in the canonical formalism that the monopole condensation causes the vanishing of the extended Polyakov loop. Therefore,
the $Z_N$ symmetry holds in the phase with the monopole condensation. 
We know that the monopole condensation leads to the quark confinement. 
Therefore, we expect that the extended Polyakov loop is an order parameter characterizing the confinement phase.
Although we have defined the extended Polyakov loop in the Minkowski space-time, 
it can be also defined in the
Euclidean space. It is interesting to see in lattice gauge theories with the condition $L>1/T$ that the extended Polyakov loop vanishes 
in  sufficiently low temperature $T$
and does not vanish in sufficiently high temperature.


 \vspace{0.2cm}
The author
expresses thanks to Prof. K. Kondo in Chiba University and 
members of theory group in KEK for useful comments
and discussions.



\end{document}